\documentclass{article}
\usepackage{spconf,amsmath,graphicx}
\usepackage{url}
\usepackage{cite}
\usepackage{subcaption}

\usepackage{multirow}
\usepackage{makecell}
\usepackage{booktabs}

\usepackage{pifont}
\newcommand{\cmark}{\ding{51}}
\newcommand{\xmark}{\ding{55}}

\def\X{{\mathbf{X}}}

\def\Y{{\mathbf{Y}}}
\def\D{{\mathbf{D}}}
\def\DTTS{{\mathbf{D}_\text{TTS}}}
\def\DASR{{\mathbf{D}_\text{ASR}}}
\def\Dtrg{{\mathbf{D}_\text{trg}}}
\def\s{{\mathbf{s}}}

\def\RefEnc{{\text{RefEnc}}}
\def\TP{{\text{TP}}}
\def\Rec{{\text{Recognizer}}}
\def\Syn{{\text{Synthesizer}}}
\def\L{{\cal L}}

\title{On Prosody Modeling for ASR+TTS based Voice Conversion}

\name{Wen-Chin Huang$^1$, Tomoki Hayashi$^1$, Xinjian Li$^2$, Shinji Watanabe$^2$, Tomoki Toda$^1$}
\address{$^1$Nagoya University, Japan\\$^2$Carnegie Mellon University, USA}

\begin{document}
\ninept
\maketitle
\begin{abstract}

In voice conversion (VC), an approach showing promising results in the latest voice conversion challenge (VCC) 2020 is to first use an automatic speech recognition (ASR) model to transcribe the source speech into the underlying linguistic contents; these are then used as input by a text-to-speech (TTS) system to generate the converted speech. Such a paradigm, referred to as ASR+TTS, overlooks the modeling of prosody, which plays an important role in speech naturalness and conversion similarity. Although some researchers have considered transferring prosodic clues from the source speech, there arises a speaker mismatch during training and conversion. To address this issue, in this work, we propose to directly predict prosody from the linguistic representation in a target-speaker-dependent manner, referred to as target text prediction (TTP). We evaluate both methods on the VCC2020 benchmark and consider different linguistic representations. The results demonstrate the effectiveness of TTP in both objective and subjective evaluations.
\end{abstract}

\section{Introduction}
\label{sec:intro}

In voice conversion (VC), one aims to convert the speech from a source to that of a target without changing the linguistic content \cite{VC, GMM-VC}. From an information perspective, the goal of VC is to extract the spoken contents from the source speech and then synthesize the converted speech from the extracted contents with the identity of the target speaker. An ideal VC system would then consist of two components, the recognition module and the synthesis module, where the two components perform the above-mentioned respective actions. Such a paradigm can be directly realized by cascading an automatic speech recognition (ASR) model and a text-to-speech (TTS) system, which we refer to as ASR+TTS. In the latest voice conversion challenge 2020 (VCC2020) \cite{vcc2020}, ASR+TTS was adopted as one of the baseline systems \cite{vcc2020-asr-tts}, and the top performing system also implemented such a framework \cite{vcc2020-task1-top}, showing state-of-the-art performance in terms of both naturalness and similarity.

Despite the promising results, the conversion and modeling of prosody in the ASR+TTS paradigm are often ignored. Prosody is a combination of several fundamental prior components in speech, such as pitch, stress, and breaks, and it impacts subsequent high-level characteristics including emotion and style. From the information perspective described in the previous paragraph, the synthesis module in ASR+TTS is responsible for recovering all information discarded by the recognition module. In the text-based ASR+TTS baseline system \cite{vcc2020-asr-tts}, since the mapping from text to prosody is one-to-many, a specific prosody or style is unpredictable from text, so the TTS model can only implicitly model the prosody pattern from the statistical properties of the training data, resulting in a collapsed, averaged prosodic style.

A closely related research field in which one aim is to control the variability of speech is expressive speech synthesis. One of the most widely used methods is the use of the global style token (GST) \cite{gst}, with which a reference speech is encoded into a fixed dimensional embedding \cite{prosody-transfer} represented as a weighted sum of a set of predefined style tokens. A TTS model equipped with GST (GST-TTS) is therefore formulated as a conditional generative model given the text, speaker identity, and prosody encoding from the reference speech. Many have generalized such a framework to variational autoencoders (VAEs) \cite{tts-vae, gmvae-tacotron} to increase the generalizability of the learnt embedding space. Another line of work focuses on fine-grained prosody control by learning variable-length prosody embeddings \cite{prosody-control-second-attention, prosody-control-forced-alignment}.

\begin{figure}[t]
	\centering
	
	\begin{subfigure}[b]{\columnwidth}
		\centering
  		\includegraphics[width=0.9\textwidth]{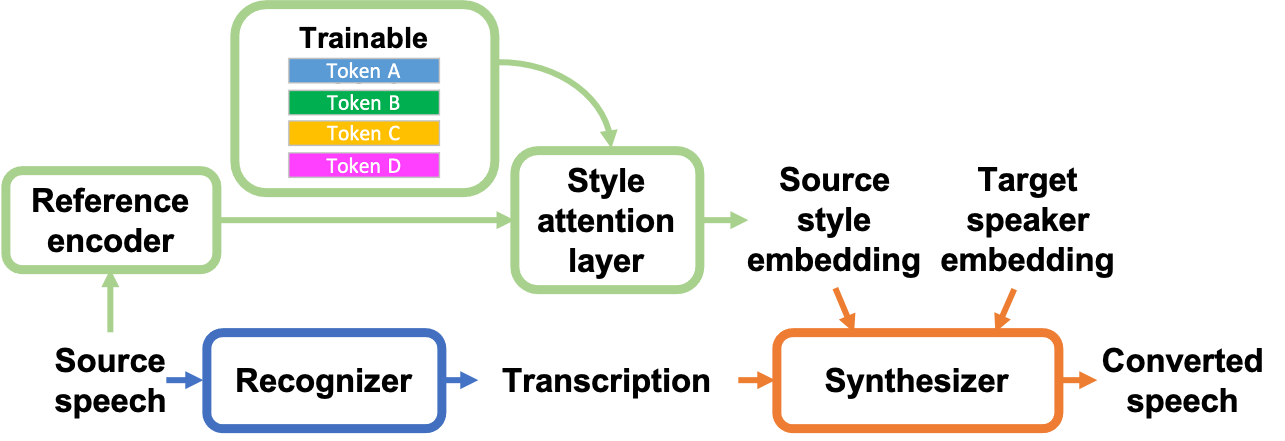}
		\caption{Source prosody trasnsfer (SPT).}
   		\label{fig:spt-conversion}
	\end{subfigure}\\
	
	\vspace{0.3cm}
	
	\begin{subfigure}[b]{\columnwidth}
		\centering
	    \includegraphics[width=0.9\textwidth]{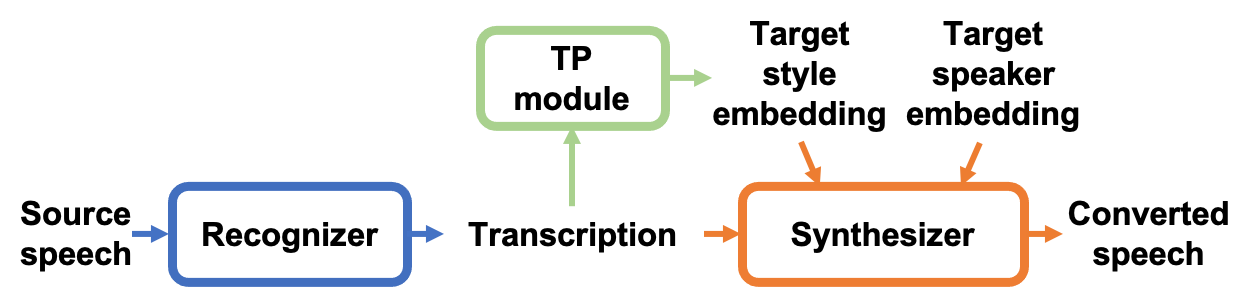}
		\caption{Target text prediction (TTP).}
   		\label{fig:ttp-conversion}
   	\end{subfigure}\\
	
	\centering
	\caption{Illustration of the conversion processes for the two prosody modeling techniques for ASR+TTS-based VC that are examined in this work.}
	\vspace{-0.3cm}
	\label{fig:prosody-modeling-conversion}
\end{figure}

\begin{figure*}[t]
	\centering
	\includegraphics[width=0.7\textwidth]{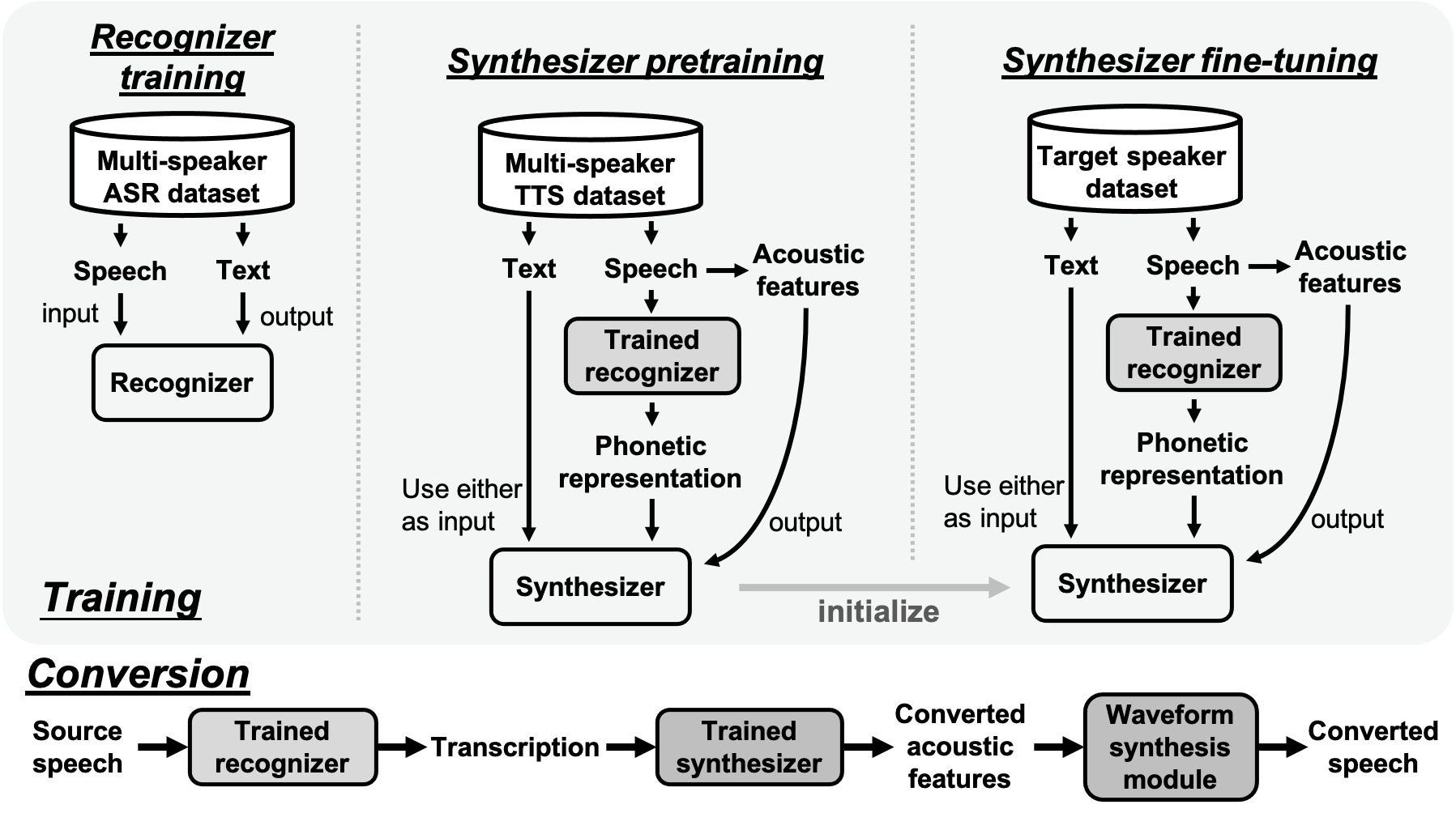} 
	\caption{Training and conversion procedures in ASR+TTS-based VC. \label{fig:asr+tts}}
	\vspace{-0.3cm}
\end{figure*}

The above-mentioned approach was first applied to VC in \cite{prosody-transfer-vc}, and it will be referred to as \textit{source prosody transfer (SPT)}. As illustrated in Figure~\ref{fig:spt-conversion}, the source speech was used as input by the reference encoder to generate a global prosody embedding such that the prosody of the converted speech follows the source. Such a process is also termed \textit{same-text prosody transfer} in the expressive TTS literature \cite{prosody-transfer} since the reference speech contains the same linguistic contents as the input of the TTS, which was also derived from the source speech. Several top performing teams in VCC2020 extended SPT by adopting a variational reference encoder augmented with a speaker adversarial layer to help with disentanglement \cite{vcc2020-task1-top, vcc2020-srcb}, showing the competitiveness of SPT.

Nonetheless, an ablation study in \cite{vcc2020-task1-top} showed that SPT did not bring about any significant improvements on task 1. We suspect that SPT can be a sub-optimal strategy for prosody modeling in ASR+TTS-based VC. First, the target-speaker-dependent TTS training causes a mismatch between training and conversion, because the speech of the target speaker is used as input to the reference encoder during training but that of the source is used during conversion. A speaker adversarial classifier can alleviate this issue \cite{vcc2020-task1-top, vcc2020-srcb} but requires careful hyperparameter tuning. Second, there are scenarios where SPT is not desired, such as emotion VC or accent conversion. 

In this work, we examine two prosody modeling methods for ASR+TTS-based VC. In addition to SPT, we propose a novel technique, which we refer to as \textit{target text prediction (TTP)}. We borrow the idea from \cite{tp-gst} and train a text prediction (TP) module to generate the prosody embedding from the text derived from the source speech. Figure~\ref{fig:ttp-conversion} illustrates this process. The TP module is first pretrained with a GST-TTS on a multispeaker dataset, and further fine-tuned in a target-speaker-dependent manner. As a result, TTP does not suffer from a mismatch between training and conversion unlike SPT. Our contributions in this work are as follows.
\begin{itemize}
	\item We propose TTP, a new prosody modeling technique for ASR+TTS-based VC, that predicts prosody in a target-speaker-dependent manner.
	\item We apply two prosody modeling methods, namely, SPT and TTP, to three ASR-TTS systems that differ in representation, and evaluate them in the two tasks in VCC2020. The results show that TTP constantly outperforms SPT.
\end{itemize}

\section{Voice conversion based on ASR+TTS}


\subsection{Overall framework and conversion process}

The ASR+TTS-based VC system in this work is built upon the baseline system for VCC2020 \cite{vcc2020-asr-tts}.
As depicted in Figure~\ref{fig:asr+tts}, the system consists of three modules: a speaker-independent recognizer, a target-speaker-dependent synthesizer, and a neural vocoder that generates the final speech waveform.
Starting from the source speech $\X$, the recognizer first extracts the spoken contents: $\hat{\Y}=\Rec(\X)$. The synthesizer takes the transcription and synthesizes the converted acoustic features: $\hat{\X} = \Syn(\hat{\Y})$\footnote{In this paper, the terms \textit{Recognizer} and \textit{Synthesizer} are used interchangeably with \textit{ASR} and \textit{TTS}, respectively}. The neural vocoder finally uses the converted acoustic features as input to reconstruct the waveform.

\subsection{Intermediate representation}

The spoken contents, $\Y$, can be any intermediate representation. In addition to using text as in \cite{vcc2020-asr-tts}, in this work we also evaluate the following two representations. Note that the goal of this work is to examine the effectiveness of the prosody modeling techniques over the three representations, but not to provide a fair comparison among them.

\noindent{\textbf{Bottleneck feature (BNF)}}: Speaker-independent, frame-level phonetic bottleneck features derived from an ASR model provide strong clues about the content. Such features were first used in \cite{VC-PPG}, where the phonetic posteriorgram (PPG), a time-versus-class matrix representing the posterior probabilities of each phonetic class (referring to a word, phone, or senone) for each frame, was used. In this work, as in \cite{S2S-iFLYTEK-VC}, we use the hidden representations before the last softmax layer in an ASR model.

\noindent{\textbf{VQW2V}}: Features such as BNF are derived from an ASR model, which requires supervision using labels; this increases the cost of building such a system, especially in low-resource or cross-lingual settings. Alternatively, several studies \cite{vqw2v-vc, vcc2020-as, fragmentvc} adopted self-supervised representations that do not require any label during training while still being speaker-independent and framewise. In this work, we adopt the vector-quantized wav2vec (VQW2V) \cite{vq-wav2vec}.

\subsection{Training}

\noindent{\textbf{ASR}}: A multispeaker dataset $\DASR$ ensures the speaker independence of the recognizer.

\noindent{\textbf{TTS}}: Synthesizer training involves a pretraining and a fine-tuning stage. Pretraining is performed on a multispeaker TTS dataset $\DTTS$, which is followed by fine-tuning on the limited target speaker dataset ${\D_{\text{trg}}}$. This is a common practice in building modern neural TTS models, as pretraining ensures stable quality and fine-tuning retains high speaker similarity \cite{semi-speech-synthesis}. Such a training strategy allows for training on even approximately 5 minutes of data. To train a text-based synthesizer, human-labeled text is used. However, since annotating ground-truth BNF and VQW2V is impossible, we use the trained recognizer to extract the representations for synthesizer training.

The ASR and TTS models adopt sequence-to-sequence (seq2seq) structures, which were shown to improve conversion similarity by modeling the long-term dependencies in speech. Note that the two models can be separately trained and thus benefit from advanced techniques and a wide variety of datasets in their own fields. 

	
	
	
	

\section{Prosody modeling in ASR+TTS-based voice conversion}

We examine two prosody modeling techniques in this work, namely, SPT and TTP. In this section, we first introduce GST and TP, which are the two fundamental building blocks, then describe the training processes in SPT and TTP. We omit the descriptions of the conversion procedures as have been discussed in Section~\ref{sec:intro}.

\subsection{Global style tokens and text prediction}

GST-TTS is aimed at using a \textit{global style embedding}\footnote{The terms such as \textit{style token} and \textit{style embedding} are terms from the original paper \cite{gst}, and readers should note they are not restricted to style but include many other factors.} (in contrast to \textit{fine-grained} embeddings in \cite{prosody-control-second-attention}) encoded from a reference speech to capture residual attributes not specified by other input streams including text, $\Y$, and speaker label, $\s$ \cite{gst}.
Formally, given a training sample $\{(\X,\Y,\s)\}$ from $\DTTS$, where $\X$ means a speech utterance, training involves minimizing the following L1-loss:
\begin{align}
    \L_\text{GST}=||\X-\Syn(\Y,\s,\text{RefEnc}(\X))||_1, \label{eq:gst_tts_loss}
\end{align}
where the $\RefEnc$ function consists of a reference encoder, a style attention layer, and a set of trainable tokens called GSTs. Specifically, the reference encoder first takes the reference speech as input and generates a fixed dimensional output vector. It is then used as the query to the attention layer to produce a set of weights over the predefined GSTs. The final style embedding is essentially the weighted sum of the GSTs and is fused with the hidden states of the TTS encoder.

During inference, the original GST-TTS requires either the reference speech or manual setting of a set of weights. To operate solely given the text input, the use of an extra TP module that takes the hidden states of the TTS encoder as input to approximate either the ground-truth style weights or the style embedding extracted from the target speech was proposed in \cite{tp-gst}. The training objective of the TP module can be formed by rewriting Equation~\ref{eq:gst_tts_loss} as
\begin{align}
    \L_\text{TP}=||\X-\Syn(\Y,\s,\TP(\Y))||_1. \label{eq:tp_loss}
\end{align}
The TP module can be trained jointly with the GST-TTS by using a stop-gradient operator. It was shown that the TP module can generate natural speech that well reflects the variations of the training data using the text input alone \cite{tp-gst}.

\subsection{Source prosody transfer}
\label{ssec:spt}

SPT was first proposed in \cite{prosody-transfer-vc} and further utilized in subsequent works \cite{vcc2020-task1-top, vcc2020-srcb, vcc2020-casia, fine-grained-prosody-vc}. The training process of SPT involves two stages.
\begin{enumerate}
    \item Pretrain GST-TTS on $\DTTS$ using $\L_\text{GST}$, as in Equation~\ref{eq:gst_tts_loss}.
    \item Finetune GST-TTS on $\Dtrg$ using $\L_\text{GST}$, as in Equation~\ref{eq:gst_tts_loss}.
\end{enumerate}
The conversion process of SPT can be formulated as
\begin{align}
    \hat{\X} = \Syn(\Rec(\X),\s_\text{trg},\RefEnc(\X)). \label{eq:spt}
\end{align}
Comparing Equation~\ref{eq:gst_tts_loss} with Equation~\ref{eq:spt}, one may find that the inputs to $\RefEnc$ are different. The speech from the target speaker is used in training, whereas the source speech is used during conversion.
As discussed in Section~\ref{sec:intro}, if $\RefEnc$ is overly fine-tuned towards the target speaker, SPT can suffer from a speaker mismatch problem. Freezing $\RefEnc$ during fine-tuning may be a remedy, but as we will show in the experiment section, this does not alleviate the problem.

\begin{table}[t]
	\centering
	\caption{Summary of the data conditions in VCC2020.}
	
	\centering
	\hspace*{-0.3cm}
	\begin{tabular}{ c | c c | c c }
		\toprule
		\multirow{2}{*}[-2pt]{Task} & \multicolumn{2}{c|}{Training phase} & \multicolumn{2}{c}{Conversion phase} \\
		\cmidrule(lr){2-5}
		& Source & Target & Source & Converted \\
		\midrule
		Task 1 & \multirow{2}{*}[-8pt]{\makecell{70 Eng.\\utterances}} & \makecell{70 Eng.\\utterances} & \multirow{2}{*}[-8pt]{\makecell{25 Eng.\\utterances}} & \multirow{2}{*}[-8pt]{\makecell{25 Eng.\\utterances}} \\
		\cmidrule(lr){1-1} \cmidrule(lr){3-3}
		Task 2 & & \makecell{70 Man./Ger./Fin.\\utterances} & & \\
		\bottomrule
	\end{tabular}
	\label{tab:data-vcc2020}
\end{table}

\subsection{Target text prediction}

We propose TTP to tackle the sub-optimal problem mentioned previously. The training process of TTP involves three stages.
\begin{enumerate}
    \item Pretrain GST-TTS on $\DTTS$.
    \item Pretrain TP on $\DTTS$ using $\L_\text{TP}$, as in Equation~\ref{eq:tp_loss} with all other model parameters fixed\footnote{This is an objective theoretically equivalent to the one in the original model \cite{tp-gst}, where the loss of the weights or embedding between those predicted from the encoder states or the speech is optimized.}. Note that the GST module is no longer required from this stage.
    \item Fine-tune TP and TTS on $\Dtrg$ using $\L_\text{TP}$, as in Equation~\ref{eq:tp_loss}.
\end{enumerate}
The conversion process of TTP can be formulated as
\begin{equation}
    \hat{\X} = \Syn(\Rec(\X),\s_\text{trg},\TP(\Rec(\X))). \label{eq:ttp}
\end{equation}
Comparing Equation~\ref{eq:tp_loss} with Equation~\ref{eq:ttp}, we see the speaker mismatch can be avoided since the TP module uses $\Y$ as input, whose speaker independence is ensured by the recognizer. TTP is therefore more robust than SPT, as we will show in the experiments.

\begin{table*}[t]
	\centering
	\caption{Objective evaluation results of task 1. Bold font indicates the best performance in the same representation. The x mark indicates no freezing of the GST-related modules, and the check mark the opposite.}
	
	\centering
	\hspace*{-0.3cm}
	\begin{tabular}{ c | c c c c | c c c c | c c c c }
		\toprule
		\multirow{2}{*}[-2pt]{System} & \multicolumn{4}{c|}{Text} & \multicolumn{4}{c|}{VQW2V} & \multicolumn{4}{c}{BNF} \\
		\cmidrule(lr){2-13}
		& MCD & F0RMSE & CER & WER & MCD & F0RMSE & CER & WER & MCD & F0RMSE & CER & WER \\
		
		\midrule
		
		Baseline & 6.49 & 29.80 & 14.4 & 24.0 & \textbf{7.13} & 31.34 & \textbf{10.5} & \textbf{17.8} & \textbf{6.86} & 30.32 & 7.9 & 13.4 \\
		\cmidrule(lr){1-13}
		SPT (\xmark) & 6.50 & 30.89 & 7.7 & 14.6 & 7.19 & \textbf{31.27} & 15.0 & 22.5 & 7.53 & 32.17 & 7.9 & 13.5 \\
		SPT (\cmark) & 6.71 & 31.82 & 7.9 & 15.1 & 7.35 & 32.46 & 13.5 & 22.6 & 7.59 & 31.92 & \textbf{7.7} & \textbf{13.2} \\ 
		\cmidrule(lr){1-13}
		TTP & \textbf{6.36} & \textbf{29.72} & \textbf{7.4} & \textbf{13.1} & \textbf{7.13} & 31.49 & 11.6 & 18.7 & 6.90 & 29.99 & \textbf{7.7} & \textbf{13.2} \\
				
		\bottomrule
	\end{tabular}
	\label{tab:obj-eval-task1}
\end{table*}

\section{Experimental settings}

\subsection{Data}

We used the VCC2020 dataset \cite{vcc2020}, which contained two tasks in our evaluation. The data conditions are summarized in Table~\ref{tab:data-vcc2020}. Both tasks share the same two source English male and female speakers whose data were not used. There were two target male and female speakers of English in task 1 whereas in task 2 there were one male and one female speaker each of Finnish, German, and Mandarin. During conversion, the source speaker's voice in the source language was converted as if it was uttered by the target speaker while keeping the linguistic contents unchanged. For each target speaker, 70 utterances in their respective languages and contents were provided. The 25 test sentences for evaluation were shared for tasks 1 and 2.

All recognizers were trained with the LibriSpeech dataset \cite{librispeech}. For the multispeaker TTS dataset, we used the ``clean'' subsets LibriTTS dataset \cite{libritts}, except in the text-based system for task 2, where we merged open-source single-speaker TTS datasets in Finnish \cite{css10}, German \cite{M-AILABS}, and Mandarin \cite{csmsc}, as in \cite{vcc2020-asr-tts}. For each of the two tasks, a separate neural vocoder was trained with the training data of the source and target speakers.

\subsection{Implementation}

The system was implemented using ESPnet, a well-developed open-source end-to-end (E2E) speech processing toolkit \cite{espnet, espnet-tts}. Following \cite{vcc2020-asr-tts}, the ASR model for the text-based system was based on the Transformer \cite{transformer, transformer-asr, transformer-asr-ctc-lm} with joint CTC/attention loss \cite{ctc-attention}, and a RNN-based language model for decoding\footnote{Code and pretrained models from the official implementation on ESPnet: \url{https://github.com/espnet/espnet/tree/master/egs/vcc20}}. The ASR model for BNF extraction was based on TDNNF-HMM \cite{tdnnf}, where we concatenated 40-dimensional MFCCs and 400-dimensional i-vectors as input. For VQW2V, we used the publicly available pretrained model provided by fairseq \cite{fairseq}\footnote{\url{https://github.com/pytorch/fairseq/tree/master/examples/wav2vec}}, as in \cite{vqw2v-vc}.

All synthesizers map their respective inputs to 80-dimensional mel filterbanks with 1024 FFT points and a 256-point frame shift (16 ms). The x-vector \cite{x-vector} was used as the speaker embedding, and we used the pretrained model provided by Kaldi \footnote{\url{https://kaldi-asr.org/models/m8}}. The average of all x-vectors of the training utterances of each speaker was used during inference. Synthesizers with discrete input including text and VQW2V had a Transformer-TTS architecture \cite{transformer-tts} with detailed settings \cite{vcc2020-asr-tts, vqw2v-vc}. For the BNF-based synthesizer, we adopted the Voice Transformer Network (VTN) \cite{VTN, VTN-TASLP} and followed the official implementation\footnote{\url{https://github.com/espnet/espnet/tree/master/egs/arctic/vc1}}. For the neural vocoder, we adopted the Parallel WaveGAN (PWG) \cite{parallel-wavegan} and followed the open-source implementation\footnote{\url{https://github.com/kan-bayashi/ParallelWaveGAN}}.

\section{Experimental evaluations}

\subsection{Objective evaluation}
\label{ssec:obj-eval}

We carried out three types of objective evaluation metric. The character/word error rates (CER/WER) from an off-the-shelf ASR system is not only an estimate of intelligibility but also a strong indicator of quality in VC, as shown in \cite{vcc2020-prediction}. Our ASR engine was Transformer-based \cite{transformer-asr} and trained on LibriSpeech. 24-dimensional mel cepstrum distortion (MCD) and the F0 root mean square error (F0RMSE) were reported to assess the spectral distortion and prosody conversion accuracy, where the calculation of both metrics were based on the WORLD vocoder \cite{WORLD}. Note that in task 2, only CER/WER were reported, since access to the ground truth speech was not available. Model selection was based on MCD and CER in tasks 1 and 2, respectively, and the best performing models generated the samples for the subjective test.

\subsubsection{Results of task 1}
\label{sssec:obj-results-task1}

Table~\ref{tab:obj-eval-task1} shows the objective results of task 1. For the text-based system, introducing prosody modeling significantly improved CER/WER. We suspect that the improved prosody pattern enabled the ASR model to easily recognize the contents correctly. Compared with SPT, TTP was shown to be more effective as it outperformed all systems. However, for frame-level features such as BNF and VQW2V, improvements from those of SPT and TTP were not significant. This result suggests that the fine-grained representations already carry abundant prosody information that is not discarded in the recognition process, such that it cannot be properly modeled by GST.

We also found that freezing the GST modules during fine-tuning degraded the performance. One possible reason is that the GSTs capture not only prosody but also other residual attributes such as channel and noise, and such mismatch must be taken care of through the fine-tuning process. Finally, compared with SPT without freezing, TTP constantly yielded a superior performance over all representations, demonstrating its superior effectiveness.

\begin{table}[t]
	\centering
	\caption{Objective evaluation results of task 2. Bold font indicates the best performance in the same representation. The x mark indicates not freezing the GST-related modules, and the check mark the opposite.}
	
	\centering
	\hspace*{-0.3cm}
	\begin{tabular}{ c c | c c | c c | c c }
		\toprule
		\multirow{2}{*}[-2pt]{System} & \multirow{2}{*}[-2pt]{Lang.} & \multicolumn{2}{c|}{Text} & \multicolumn{2}{c|}{VQW2V} & \multicolumn{2}{c}{BNF} \\
		\cmidrule(lr){3-8}
		& & CER & WER & CER & WER & CER & WER \\
		
		\midrule
		
		\multirow{4}{*}[-2pt]{Baseline} & Fin. & 39.1 & 67.1 & 12.6 & 21.9 & 9.4 & 16.1 \\
		& Ger. & 11.0 & 18.8 & 9.3 & 15.5 & 7.8 & 13.6 \\
		& Man. & 13.3 & 23.1 & 8.2 & 16.1 & 8.4 & 14.0 \\
		\cmidrule(lr){2-8}
		& Avg. & \textbf{21.1} & \textbf{36.3} & 10.0 & 17.8 & 8.5 & 14.5 \\
		
		\cmidrule(lr){1-8}
		
		\multirow{4}{*}[-2pt]{SPT (\xmark)} & Fin. & 47.8 & 79.9 & 12.5 & 19.9 & 8.7 & 14.6 \\
		& Ger. & 7.9 & 14.7 & 9.8 & 17.7 & 9.2 & 15.8 \\
		& Man. & 8.6 & 15.3 & 8.3 & 16.6 & 8.6 & 14.8 \\
		\cmidrule(lr){2-8}
		& Avg. & 21.4 & 36.6 & 10.2 & 18.0 & 8.8 & 15.0 \\
				
		\cmidrule(lr){1-8}
		
		\multirow{4}{*}[-2pt]{SPT (\cmark)} & Fin. & 47.4 & 80.7 & 12.3 & 21.1 & 8.9 & 15.3 \\
		& Ger. & 8.6 & 15.9 & 10.2 & 19.0 & 8.6 & 15.4 \\
		& Man. & 8.3 & 15.8 & 8.2 & 15.1 & 8.1 & 14.2 \\
		\cmidrule(lr){2-8}
		& Avg. & 21.4 & 37.4 & 10.2 & 18.4 & 8.5 & 14.9 \\ 

		\cmidrule(lr){1-8}		
		
		\multirow{4}{*}[-2pt]{TTP} & Fin. & 51.6 & 80.5 & 10.8 & 19.9 & 8.9 & 15.0 \\
		& Ger. & 11.3 & 20.4 & 9.9 & 17.2 & 8.0 & 13.9 \\
		& Man. & 8.8 & 17.1 & 7.9 & 15.3 & 7.9 & 13.6 \\
		\cmidrule(lr){2-8}
		& Avg. & 23.9 & 39.3 & \textbf{9.5} & \textbf{17.5} & \textbf{8.2} & \textbf{14.2} \\
				
		\bottomrule
	\end{tabular}
	\label{tab:obj-eval-task2}
\end{table}

\subsubsection{Results of task 2}
\label{sssec:obj-results-task2}

Table~\ref{tab:obj-eval-task2} shows the objective results of task 2. First, for the text-based system, SPT improved the performance for German and Mandarin target speakers, and TTP was beneficial for Mandarin speakers. By listening to several samples, we found that both SPT and TTP helped the model insert proper short pauses, which significantly increased intelligibility. On the other hand, both methods degraded the performance for the Finnish speakers; we assume that the improper text preprocessing for Finnish \cite{vcc2020-asr-tts} made both acoustic and prosody modeling difficult. For the framewise representations, we conclude that neither SPT nor TTP much affected the intelligibility. Note that freezing GST in SPT did not cause degradation as it did in task 1.

\begin{figure}[t]
	\centering
	
	\begin{subfigure}[b]{\columnwidth}
		\centering
  		\includegraphics[width=\textwidth]{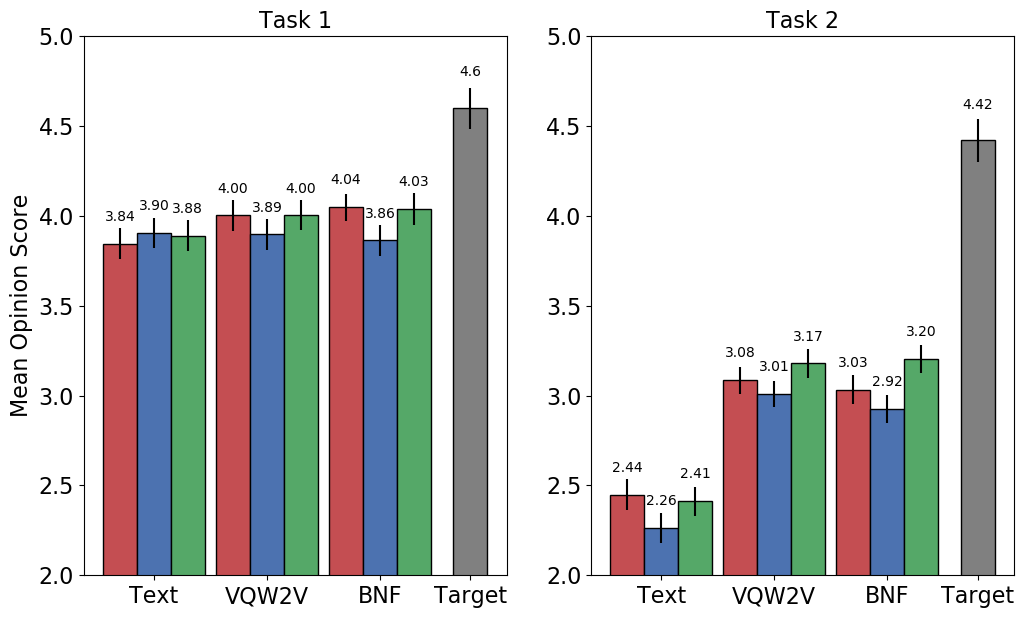}
	\end{subfigure}\\
	
	\begin{subfigure}[b]{\columnwidth}
		\centering
  		\includegraphics[width=\textwidth]{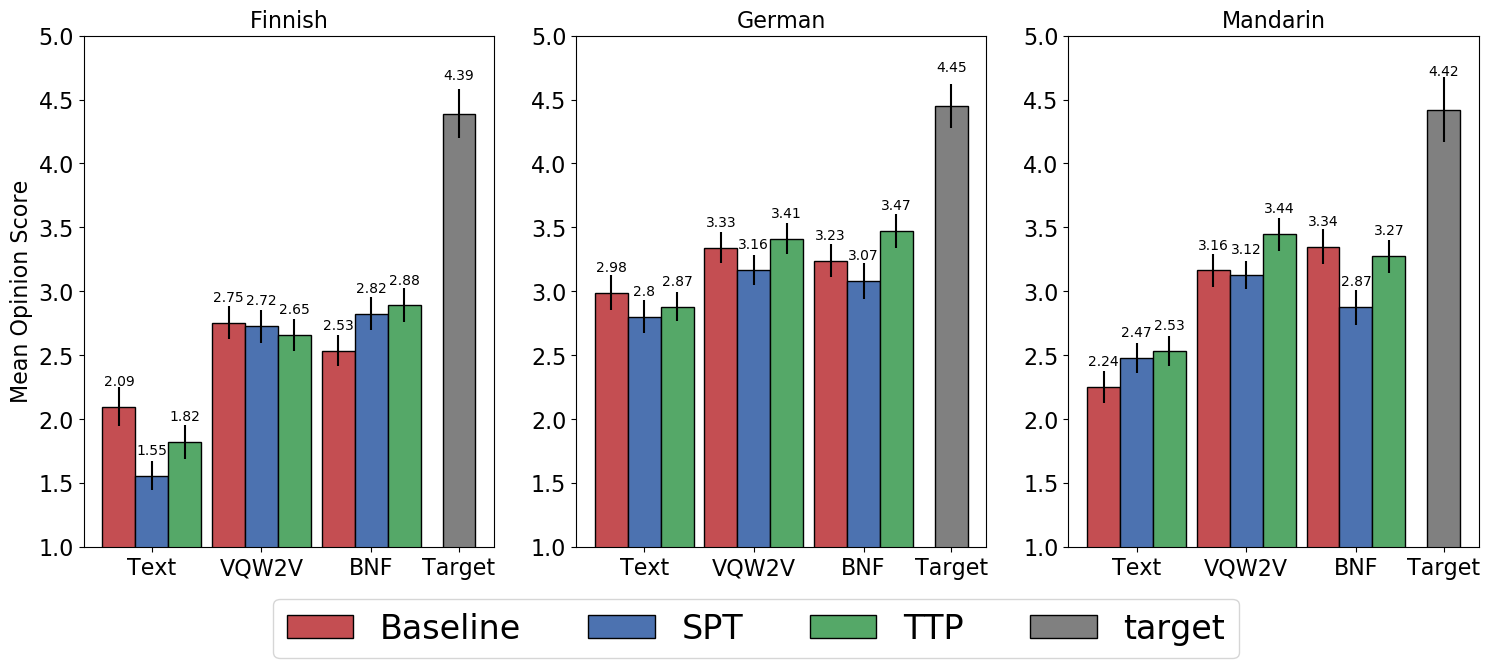}
	\end{subfigure}\\
	
	\centering
	\caption{Naturalness MOS plots. Top row: taskwise results. Bottom row: Language breakdown for results of task 2.}
	\label{fig:mos}
\end{figure}

\begin{figure}[t]
	\centering
	\includegraphics[width=\columnwidth]{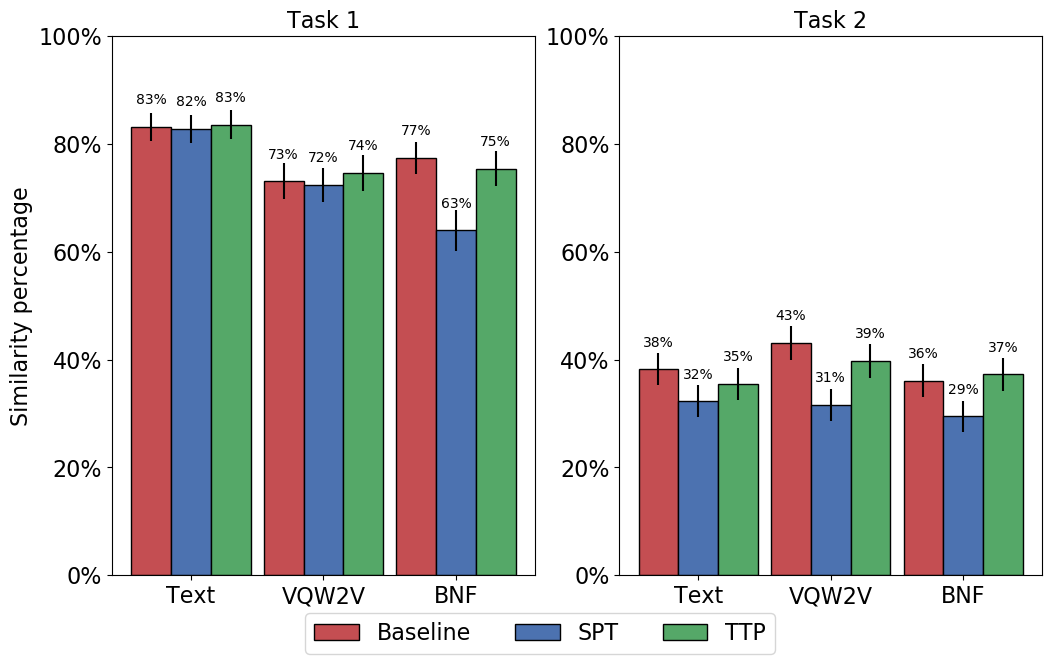}	
	
	\caption{Taskwise similarity results.}
	\label{fig:sim}
\end{figure}

\subsection{Subjective evaluation}

We followed the subjective evaluation methodology described in \cite{vcc2020} and evaluated two aspects: naturalness and speaker similarity.
For naturalness, participants were asked to evaluate the naturalness of the speech by the mean opinion score (MOS) test on a five-point scale.
For conversion similarity, each listener was presented a natural target speech and a converted speech and asked to judge whether they were produced by the same speaker on a four-point scale.

For each system, five random utterances were chosen for each conversion pair. In the naturalness test, recordings of the target speakers were also included and served as the upper bound. In the similarity test for task 2, following \cite{vcc2020}, we selected three English recordings and two L2 language recordings as the natural reference for the five converted utterances.
All subjective evaluations were performed using the open-source toolkit \cite{p808-open-source} that implements the ITU-T Recommendation P.808 \cite{p808} for subjective speech quality assessment in a crowd using the Amazon Mechanical Turk (Mturk) and screens the obtained data for unreliable ratings. We recruited more than 100 listeners from the United States and had each sample rated by five different participants on average.
Note that to reduce cost, we eliminated SPT systems that freeze GST in the listening tests, since they yield inferior performance compared with systems that do not freeze GST, as shown in Section~\ref{ssec:obj-eval}.
Audio samples are available online\footnote{\url{https://unilight.github.io/Publication-Demos/publications/prosody-asr-tts-vc/index.html}}.

\subsubsection{Naturalness test}
\label{sssec:naturalness-results}

Figure~\ref{fig:mos} shows the naturalness results. We focus on task 1 first. For text-based systems, contrary to the findings in Section~\ref{sssec:obj-results-task1}, SPT and TTP results were comparable to the baseline. For frame-level features, SPT largely degraded the performance, whereas TTP could compensate for the degradation but only to make it on par with the baseline. We can still conclude that TTP was superior to SPT for task 1, but no significant improvements were observed when compared with the baseline. These results are somewhat consistent with the findings in \cite{vcc2020-task1-top}.

For task 2, all systems showed performance degradation brought about by SPT, and TTP significantly outperformed SPT for all representations. For systems based on frame-level features, TTP could even outperform the baseline and was also comparable to the text-based system. To investigate this gap, we plotted the breakdown per target language in the bottom graphs of Figure~\ref{fig:mos}. We suspect that the relative performance change was again correlated with the text preprocessing, as stated in Section~\ref{sssec:obj-results-task2}. As in \cite{vcc2020-asr-tts}, thanks to the open-source community, we utilized G2P tools to convert both English and Mandarin text into phonemes, resulting in a better acoustic model and a larger improvement brought about by TTP. On the other hand, because of the lack of linguistic knowledge, characters were used for Finnish and German, resulting in degradation when combined with TTP. We thus conclude that TTP is an effective method for improving naturalness in task 2, if the input representation is properly processed.

\subsubsection{Similarity test}
\label{sssec:similarity-test}

Figure~\ref{fig:sim} shows the similarity results. In task 1, both SPT and TTP demonstrated a comparable performance for the text and VQW2V-based systems, whereas SPT performed significantly worse in the BNF-based system. As for task 2, all three representations yielded a similar trend such that no obvious performance gain could be distinguished for any prosody modeling method. Although it it widely believed that prosody is highly correlated to speaker identity, here we could only conclude that prosody modeling only improved naturalness. 

On the other hand, for situations where the L2 recordings were used as the reference speech, it was necessary to ``imagine'' the target speaker's voice in English in order to judge the similarity to the converted speech. Under such a subjective evaluation protocol, we suspect that listeners could only resort to coarse prosodic clues such as the overall pitch level and timbre, which were not affected by the prosody modeling techniques.

\begin{figure}[t]
	\centering
	\begin{subfigure}{.5\columnwidth}
	    \centering
	    \includegraphics[width=\textwidth]{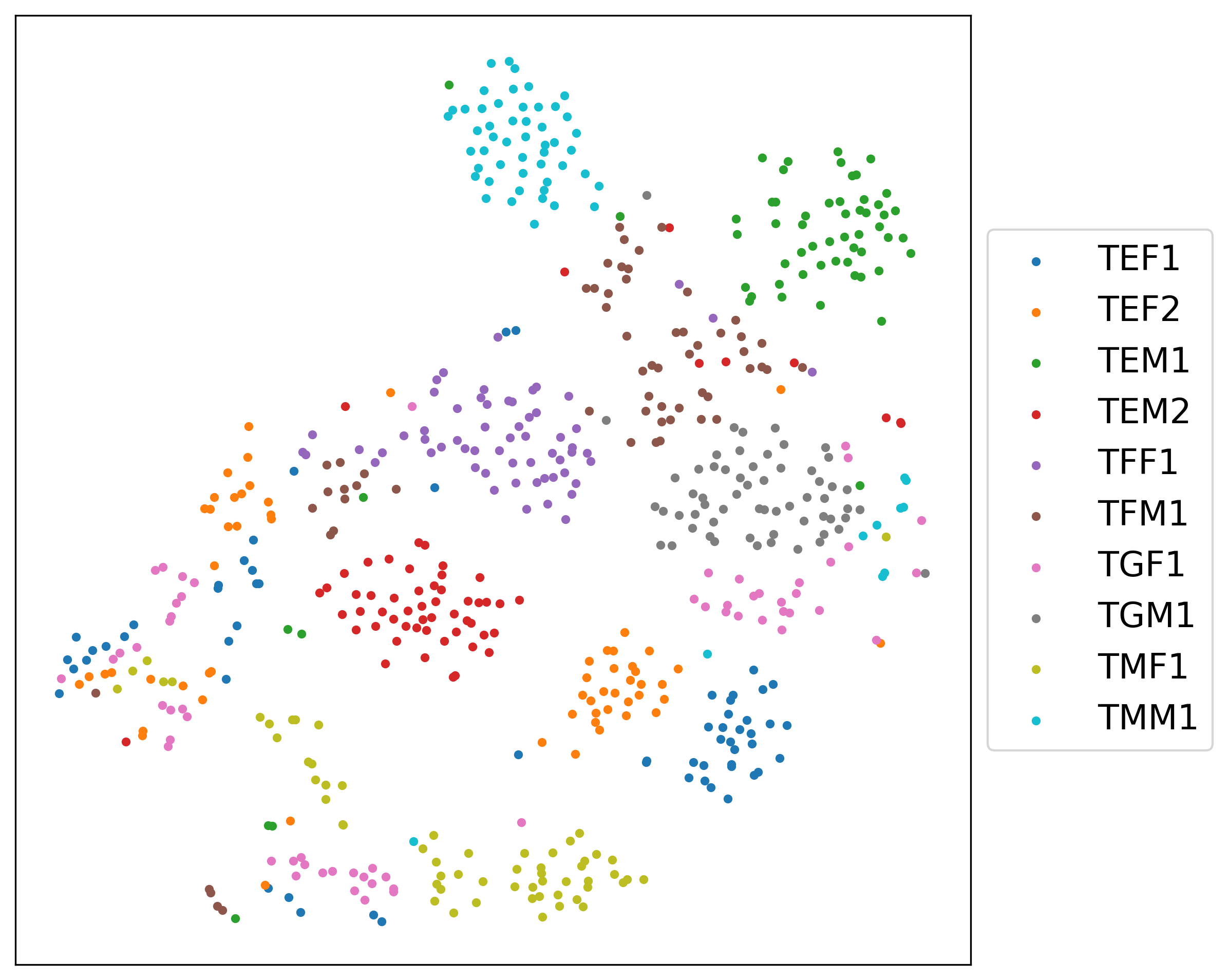}
	    \caption{Text}
   		\label{fig:tsne_text}
	\end{subfigure}%
	\begin{subfigure}{.5\columnwidth}
	    \centering
	    \includegraphics[width=\textwidth]{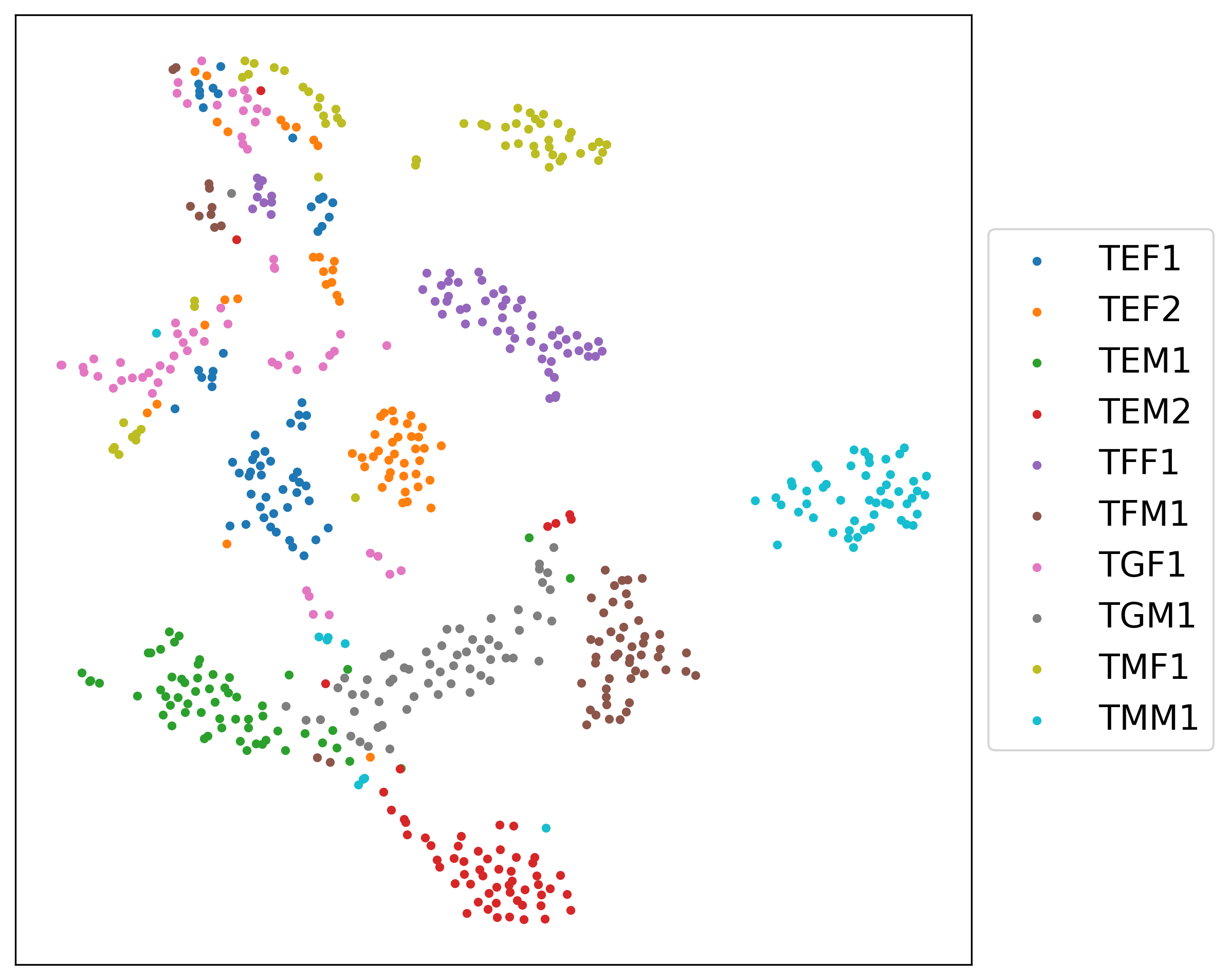}
	    \caption{VQW2V}
   		\label{fig:tsne_vqw2v}
	\end{subfigure}
	
	\begin{subfigure}{.5\columnwidth}
	    \centering
	    \includegraphics[width=\textwidth]{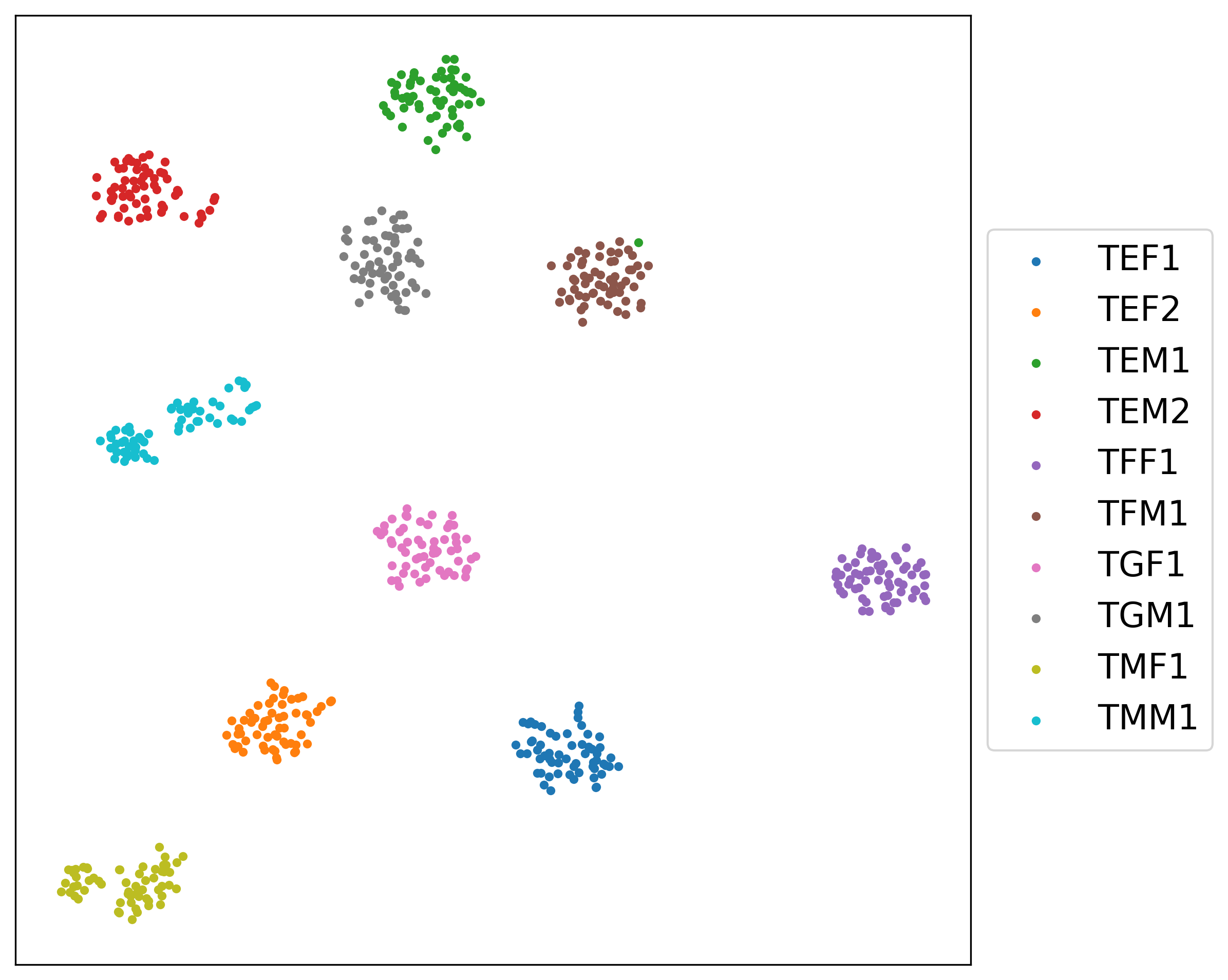}
	    \caption{BNF}
   		\label{fig:tsne_bnf}
	\end{subfigure}%
	\begin{subfigure}{.5\columnwidth}
	    \centering
	    \includegraphics[width=\textwidth]{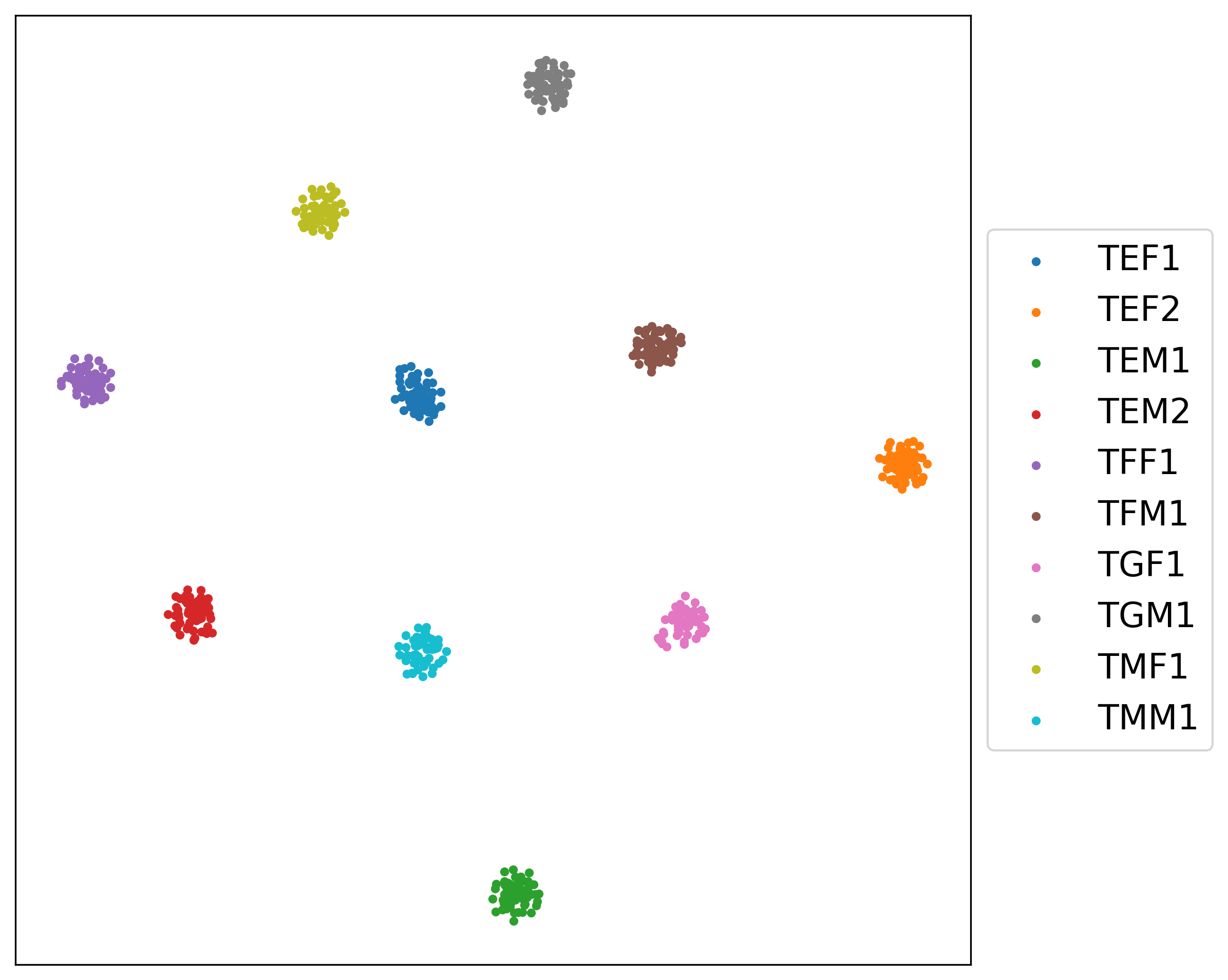}
	    \caption{X-vector}
   		\label{fig:tsne_x-vector}
	\end{subfigure}

	\centering
	\caption{Visualizations of style embeddings and X-vectors of training utterances of the 10 target speakers in VCC2020. Style embeddings were extracted using GST-TTS models pretrained on LibriTTS using different input representations. Each dot represents the embedding of one utterance.}
	\label{fig:tsne}
\end{figure}

\subsection{Visualization of style embeddings}

We visualized the style embedding space of the GST-TTS models using different representations that are pretrained on LibriTTS (which is the resulting model of stage 1 in Section~\ref{ssec:spt}). We input the training utterances of the 10 target speakers to obtain the style embeddings, and performed dimensional reduction by the t-SNE method \cite{tsne} for visualization. As a reference, we also visualized the X-vectors. Finally, we colored the dots with respect to each target speaker. The resulting plots are depicted in Figure~\ref{fig:tsne}.

First, it can be observed that all representations (text, VQW2V, and BNF) were somehow clustered with respect to speakers. As one possible reason may be that prosody is speaker-dependent \cite{capacitron}, we may also infer that the use of X-vectors for speaker embeddings cannot fully capture all speaker-related variations, such that the GST-TTS model relies on the help of the style embedding. This justifies the use of the speaker adversarial classifier \cite{vcc2020-task1-top, vcc2020-srcb}.

We further observed that the order of visual \textit{degree of clusterness} (how close together the dots of a speaker are) from low to high is text, VQW2V, BNF, and X-vector. We suggest that the degree of clusterness reflects the amount of variation that style embeddings can explain. As we know that text contains the least prosody information, we could infer that owing to the frame-level nature, representations such as VQW2V and BNF already contained much prosody information such that there was not much left for the style embeddings to capture. Nonetheless, from the improvements of naturalness in task 2, as described in Section~\ref{sssec:naturalness-results}, we note that prosody modeling is still effective in capturing residual information.

\section{Conclusions}

We examined two prosody modeling methods, namely, SPT and TTP, for ASR+TTS-based VC. Whereas SPT had already been applied by several top systems in VCC2020, TTP was newly proposed in this work, with the motivation to prevent the mismatch between training and conversion in SPT. We conducted experiments on the VCC2020 benchmark and considered three intermediate representations: text, BNF, and VQW2V. Results showed that TTP was consistently superior to SPT and could outperform the baseline with proper input representation. Finally, the visualization of the style embeddings learnt with different representations shed light on the limited improvements brought about by SPT and TTP.

\section{Acknowledgements}

This work was partly supported by JSPS KAKENHI Grant Number 21J20920 and JST CREST Grant Number JPMJCR19A3, Japan. We would also like to thank Yu-Huai Peng and Hung-Shin Lee from Academia Sinica, Taiwan, for training the BNF extractor.

\bibliographystyle{IEEEbib}
\bibliography{ref}

\begin{thebibliography}{10}

\bibitem{VC}
Y.~Stylianou, O.~Cappe, and E.~Moulines,
\newblock ``{Continuous probabilistic transform for voice conversion},''
\newblock {\em IEEE TSAP}, vol. 6, no. 2, pp. 131--142, 1998.

\bibitem{GMM-VC}
T.~Toda, A.~W. Black, and K.~Tokuda,
\newblock ``{Voice Conversion Based on Maximum-Likelihood Estimation of
  Spectral Parameter Trajectory},''
\newblock {\em IEEE TASLP}, vol. 15, no. 8, pp. 2222--2235, 2007.

\bibitem{vcc2020}
Y.~Zhao, W.-C. Huang, X.~Tian, J.~Yamagishi, R.~K. Das, T.~Kinnunen, Z.~Ling,
  and T.~Toda,
\newblock ``{Voice Conversion Challenge 2020 - Intra-lingual semi-parallel and
  cross-lingual voice conversion -},''
\newblock in {\em Proc. Joint Workshop for the BC and VCC 2020}, 2020, pp.
  80--98.

\bibitem{vcc2020-asr-tts}
W.-C. Huang, T.~Hayashi, S.~Watanabe, and T.~Toda,
\newblock ``{The Sequence-to-Sequence Baseline for the Voice Conversion
  Challenge 2020: Cascading ASR and TTS},''
\newblock in {\em Proc. Joint Workshop for the BC and VCC 2020}, 2020, pp.
  160--164.

\bibitem{vcc2020-task1-top}
J.-X. Zhang, L.-J. Liu, Y.-N. Chen, Y.-J. Hu, Y.~J., Z.-H. Ling, and L.-R. Dai,
\newblock ``{Voice Conversion by Cascading Automatic Speech Recognition and
  Text-to-Speech Synthesis with Prosody Transfer},''
\newblock in {\em Proc. Joint Workshop for the BC and VCC 2020}, 2020, pp.
  121--125.

\bibitem{gst}
Y.~Wang, D.~Stanton, Y.~Zhang, RJ-Skerry Ryan, E.~Battenberg, J.~Shor, Y.~Xiao,
  Y.~Jia, F.~Ren, and R.~A. Saurous,
\newblock ``{Style Tokens: Unsupervised Style Modeling, Control and Transfer in
  End-to-End Speech Synthesis},''
\newblock in {\em Proc. ICML}, 2018, pp. 5180--5189.

\bibitem{prosody-transfer}
RJ~Skerry-Ryan, E.~Battenberg, Y.~Xiao, Y.~Wang, D.~Stanton, J.~Shor, R.~Weiss,
  R.~Clark, and R.~A. Saurous,
\newblock ``{Towards End-to-End Prosody Transfer for Expressive Speech
  Synthesis with Tacotron},''
\newblock in {\em Proc. ICML}, 2018, pp. 4693--4702.

\bibitem{tts-vae}
Y.-J. Zhang, S.~Pan, L.~He, and Z.-H. Ling,
\newblock ``{Learning Latent Representations for Style Control and Transfer in
  End-to-end Speech Synthesis},''
\newblock in {\em Proc. ICASSP}, 2019, pp. 6945--6949.

\bibitem{gmvae-tacotron}
W.-N. Hsu, Y.~Zhang, R.~Weiss, H.~Zen, Y.~Wu, Y.~Cao, and Y.~Wang,
\newblock ``{Hierarchical Generative Modeling for Controllable Speech
  Synthesis},''
\newblock in {\em Proc. ICLR}, 2019.

\bibitem{prosody-control-second-attention}
Y.~Lee and T.~Kim,
\newblock ``{Robust and Fine-grained Prosody Control of End-to-end Speech
  Synthesis},''
\newblock in {\em Proc. ICASSP}, 2019, pp. 5911--5915.

\bibitem{prosody-control-forced-alignment}
V.~Klimkov, S.~Ronanki, J.~Rohnke, and T.~Drugman,
\newblock ``{Fine-Grained Robust Prosody Transfer for Single-Speaker Neural
  Text-To-Speech},''
\newblock in {\em Proc. Interspeech}, 2019, pp. 4440--4444.

\bibitem{prosody-transfer-vc}
S.~Liu, Y.~Cao, S.~Kang, N.~Hu, X.~Liu, D.~Su, D.~Yu, and H.~Meng,
\newblock ``{Transferring Source Style in Non-Parallel Voice Conversion},''
\newblock in {\em Proc. Interspeech}, 2020, pp. 4721--4725.

\bibitem{vcc2020-srcb}
Q.~Ma, R.~Liu, X.~Wen, C.~Lu, and X.~Chen,
\newblock ``{Submission from SRCB for Voice Conversion Challenge 2020},''
\newblock in {\em Proc. Joint Workshop for the BC and VCC 2020}, 2020, pp.
  131--135.

\bibitem{tp-gst}
D.~Stanton, Y.~Wang, and RJ~Skerry-Ryan,
\newblock ``{Predicting Expressive Speaking Style from Text in End-To-End
  Speech Synthesis},''
\newblock in {\em Proc. SLT}, 2018, pp. 595--602.

\bibitem{VC-PPG}
L.~Sun, K.~Li, H.~Wang, S.~Kang, and H.~Meng,
\newblock ``{Phonetic posteriorgrams for many-to-one voice conversion without
  parallel data training},''
\newblock in {\em Proc. ICME}, 2016, pp. 1--6.

\bibitem{S2S-iFLYTEK-VC}
J.~{Zhang}, Z.~{Ling}, L.~{Liu}, Y.~{Jiang}, and L.~{Dai},
\newblock ``{Sequence-to-Sequence Acoustic Modeling for Voice Conversion},''
\newblock {\em IEEE/ACM TASLP}, vol. 27, no. 3, pp. 631--644, 2019.

\bibitem{vqw2v-vc}
W.-C. Huang, Y.-C. Wu, T.~Hayashi, and T.~Toda,
\newblock ``{Any-to-One Sequence-to-Sequence Voice Conversion using
  Self-Supervised Discrete Speech Representations},''
\newblock in {\em Proc. ICASSP}, 2021, pp. 5944--5948.

\bibitem{vcc2020-as}
Y.-H. Peng, C.-H. Hu, A.~Kang, H.-S. Lee, P.-Y. Chen, Y.~Tsao, and H.-M. Wang,
\newblock ``{The Academia Sinica Systems of Voice Conversion for VCC2020},''
\newblock in {\em Proc. Joint Workshop for the BC and VCC 2020}, 2020, pp.
  180--183.

\bibitem{fragmentvc}
Y.~Y Lin, C.-M. Chien, J.-H. Lin, H.-Y. Lee, and L.-S. Lee,
\newblock ``{FragmentVC: Any-to-Any Voice Conversion by End-to-End Extracting
  and Fusing Fine-Grained Voice Fragments With Attention},''
\newblock in {\em Proc. ICASSP}, 2021, pp. 5939--5943.

\bibitem{vq-wav2vec}
A.~Baevski, S.~Schneider, and M.~Auli,
\newblock ``{vq-wav2vec: Self-Supervised Learning of Discrete Speech
  Representations},''
\newblock in {\em Proc. ICLR}, 2020.

\bibitem{semi-speech-synthesis}
K.~{Inoue}, S.~{Hara}, M.~{Abe}, T.~{Hayashi}, R.~{Yamamoto}, and
  S.~{Watanabe},
\newblock ``{Semi-Supervised Speaker Adaptation for End-to-End Speech Synthesis
  with Pretrained Models},''
\newblock in {\em Proc. ICASSP}, 2020, pp. 7634--7638.

\bibitem{vcc2020-casia}
L.~Zheng, J.~Tao, Z.~Wen, and R.~Zhong,
\newblock ``{CASIA Voice Conversion System for the Voice Conversion Challenge
  2020},''
\newblock in {\em Proc. Joint Workshop for the BC and VCC 2020}, 2020, pp.
  136--139.

\bibitem{fine-grained-prosody-vc}
Z.~Lian, R.~Zhong, Z.~Wen, B.~Liu, and J.~Tao,
\newblock ``{Towards Fine-Grained Prosody Control for Voice Conversion},''
\newblock in {\em Proc. ISCSLP}, 2021, pp. 1--5.

\bibitem{librispeech}
V.~{Panayotov}, G.~{Chen}, D.~{Povey}, and S.~{Khudanpur},
\newblock ``{LibriSpeech: An ASR corpus based on public domain audio books},''
\newblock in {\em Proc. ICASSP}, 2015, pp. 5206--5210.

\bibitem{libritts}
H.~Zen, V.~Dang, R.~Clark, Y.~Zhang, R.~J. Weiss, Y.~Jia, Z.~Chen, and Y.~Wu,
\newblock ``{LibriTTS: A Corpus Derived from LibriSpeech for Text-to-Speech},''
\newblock in {\em Proc. Interspeech}, 2019, pp. 1526--1530.

\bibitem{css10}
K.~Park and T.~Mulc,
\newblock ``{CSS10: A Collection of Single Speaker Speech Datasets for 10
  Languages},''
\newblock in {\em Proc. Interspeech}, 2019, pp. 1566--1570.

\bibitem{M-AILABS}
{Munich Artificial Intelligence Laboratories GmbH},
\newblock ``The {M-AILABS} speech dataset,'' 2019,
\newblock accessed 30 November 2019.

\bibitem{csmsc}
{Data Baker China},
\newblock ``Chinese standard mandarin speech corpus,'' accessed 05 May 2020.

\bibitem{espnet}
S.~Watanabe, T.~Hori, S.~Karita, T.~Hayashi, J.~Nishitoba, Y.~Unno, N.~{E. Y.
  Soplin}, J.~Heymann, M.~Wiesner, N.~Chen, A.~Renduchintala, and T.~Ochiai,
\newblock ``{ESPnet: End-to-End Speech Processing Toolkit},''
\newblock in {\em Proc. Interspeech}, 2018, pp. 2207--2211.

\bibitem{espnet-tts}
T.~{Hayashi}, R.~{Yamamoto}, K.~{Inoue}, T.~{Yoshimura}, S.~{Watanabe},
  T.~{Toda}, K.~{Takeda}, Y.~{Zhang}, and X.~{Tan},
\newblock ``{Espnet-TTS: Unified, Reproducible, and Integratable Open Source
  End-to-End Text-to-Speech Toolkit},''
\newblock in {\em Proc. ICASSP}, 2020, pp. 7654--7658.

\bibitem{transformer}
A.~Vaswani, N.~Shazeer, N.~Parmar, J.~Uszkoreit, L.~Jones, A.~N Gomez,
  L.~Kaiser, and I.~Polosukhin,
\newblock ``{Attention is All you Need},''
\newblock in {\em Proc. NIPS}, 2017, pp. 5998--6008.

\bibitem{transformer-asr}
L.~{Dong}, S.~{Xu}, and B.~{Xu},
\newblock ``{Speech-Transformer: A No-Recurrence Sequence-to-Sequence Model for
  Speech Recognition},''
\newblock in {\em Proc. ICASSP}, 2018, pp. 5884--5888.

\bibitem{transformer-asr-ctc-lm}
S.~Karita, N.~E.~Y. Soplin, S.~Watanabe, M.~Delcroix, A.~Ogawa, and
  T.~Nakatani,
\newblock ``{Improving Transformer-Based End-to-End Speech Recognition with
  Connectionist Temporal Classification and Language Model Integration},''
\newblock in {\em Proc. Interspeech}, 2019, pp. 1408--1412.

\bibitem{ctc-attention}
S.~{Watanabe}, T.~{Hori}, S.~{Kim}, J.~R. {Hershey}, and T.~{Hayashi},
\newblock ``{Hybrid CTC/Attention Architecture for End-to-End Speech
  Recognition},''
\newblock {\em IEEE Jour. of Selected Topics in Signal Processing}, vol. 11,
  no. 8, pp. 1240--1253, 2017.

\bibitem{tdnnf}
D.~Povey, G.~Cheng, Y.~Wang, K.~Li, H.~Xu, M.~Yarmohammadi, and S.~Khudanpur,
\newblock ``{Semi-Orthogonal Low-Rank Matrix Factorization for Deep Neural
  Networks},''
\newblock in {\em Proc. Interspeech}, 2018, pp. 3743--3747.

\bibitem{fairseq}
M.~Ott, S.~Edunov, A.~Baevski, A.~Fan, S.~Gross, N.~Ng, D.~Grangier, and
  M.~Auli,
\newblock ``{fairseq: A Fast, Extensible Toolkit for Sequence Modeling},''
\newblock in {\em Proc. NAACL}, 2019, pp. 48--53.

\bibitem{x-vector}
D.~Snyder, D.~Garcia-Romero, G.~Sell, D.~Povey, and S.~Khudanpur,
\newblock ``X-vectors: Robust dnn embeddings for speaker recognition,''
\newblock in {\em Proc. ICASSP}, 2018, pp. 5329--5333.

\bibitem{transformer-tts}
N.~Li, S.~Liu, Y.~Liu, S.~Zhao, and M.~Liu,
\newblock ``{Neural Speech Synthesis with Transformer Network},''
\newblock in {\em Proc. AAAI}, 2019, pp. 6706--6713.

\bibitem{VTN}
W.-C. Huang, T.~Hayashi, Y.-C. Wu, H.~Kameoka, and T.~Toda,
\newblock ``{Voice Transformer Network: Sequence-to-Sequence Voice Conversion
  Using Transformer with Text-to-Speech Pretraining},''
\newblock in {\em Proc. Interspeech}, 2020, pp. 4676--4680.

\bibitem{VTN-TASLP}
W.-C. Huang, T.~Hayashi, Y.-C. Wu, H.~Kameoka, and T.~Toda,
\newblock ``{Pretraining Techniques for Sequence-to-Sequence Voice
  Conversion},''
\newblock {\em IEEE TASLP}, vol. 29, pp. 745--755, 2021.

\bibitem{parallel-wavegan}
R.~{Yamamoto}, E.~{Song}, and J.~{Kim},
\newblock ``{Parallel WaveGAN: A Fast Waveform Generation Model Based on
  Generative Adversarial Networks with Multi-Resolution Spectrogram},''
\newblock in {\em Proc. ICASSP}, 2020, pp. 6199--6203.

\bibitem{vcc2020-prediction}
R.~Kumar Das, T.~Kinnunen, W.-C. Huang, Z.-H. Ling, J.~Yamagishi, Z.~Yi,
  X.~Tian, and T.~Toda,
\newblock ``{Predictions of Subjective Ratings and Spoofing Assessments of
  Voice Conversion Challenge 2020 Submissions},''
\newblock in {\em Proc. Joint Workshop for the BC and VCC 2020}, 2020, pp.
  99--120.

\bibitem{WORLD}
M.~{Morise}, F.~{Yokomori}, and K.~{Ozawa},
\newblock ``{WORLD: A Vocoder-Based High-Quality Speech Synthesis System for
  Real-Time Applications},''
\newblock {\em IEICE Transactions on Information and Systems}, vol. 99, pp.
  1877--1884, 2016.

\bibitem{p808-open-source}
B.~Naderi and R.~Cutler,
\newblock ``{An Open Source Implementation of ITU-T Recommendation P.808 with
  Validation},''
\newblock in {\em Proc. Interspeech}, 2020, pp. 2862--2866.

\bibitem{p808}
{ITU-T Recommendation P.808},
\newblock ``{Subjective evaluation of speech quality with a crowdsourcing
  approach},'' 2018.

\bibitem{tsne}
L.~van~der Maaten and G.~Hinton,
\newblock ``{Visualizing data using t-SNE},''
\newblock {\em JMLR}, vol. 9, pp. 2579--2605, 2008.

\bibitem{capacitron}
E.~Battenberg, S.~Mariooryad, D.~Stanton, RJ~Skerry-Ryan, M.~Shannon, D.~Kao,
  and T.~Bagby,
\newblock ``{Effective use of variational embedding capacity in expressive
  end-to-end speech synthesis},''
\newblock {\em arXiv preprint arXiv:1906.03402}, 2019.

\end{thebibliography}

\end{document}